%% file: multi.tex
\documentclass[12pt,a4paper]{conference}

\usepackage{fancyhdr}
\usepackage{graphicx,amsmath,amssymb,cite}
\usepackage{multind}
\makeindex{author} \makeindex{subject}

\input MENU07macros.tex

\begin{document}

\Chapter{\boldmath MULTIMESON PRODUCTION IN $pp$ INTERACTIONS AS A BACKGROUND 
FOR $\eta$ AND $\eta'$ DECAYS}
           {Multimeson Production}{A. Kupsc \it{et al.}}
\vspace{-6 cm}\includegraphics[width=6 cm]{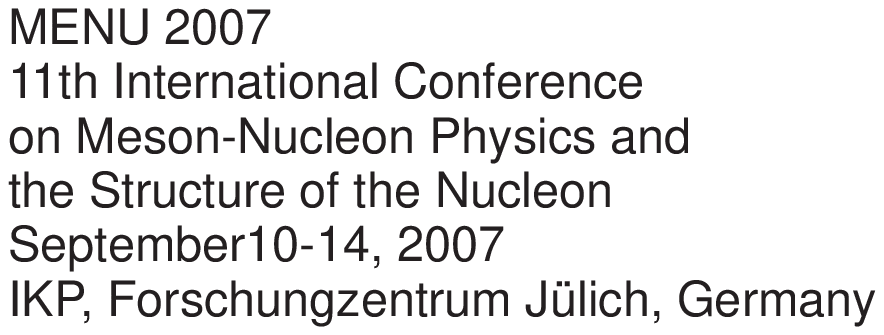}
\vspace{4 cm}

\addcontentsline{toc}{chapter}{{\it  A. Kupsc}} \label{authorStart}

\newcommand{\et}{\mbox{$\eta$}}
\newcommand{\pppipipi}{\mbox{$pp\to pp\pi^{+}\pi^{-}\po$}}
\newcommand{\pppopopo}{\mbox{$pp\to pp\po\po\po$}}
\newcommand{\pptpi}{\mbox{$pp\to pp \pi\pi\pi$}}
\newcommand{\etp}{\mbox{$\eta'$}}
\newcommand{\po}{\mbox{$\pi^0$}}
\newcommand{\gm}{\mbox{$\gamma$}}
\newcommand{\pim}{\mbox{$\pi^-$}}
\newcommand{\pip}{\mbox{$\pi^+$}}
\newcommand{\elp}{\mbox{$e^+$}}
\newcommand{\elm}{\mbox{$e^-$}}
\newcommand{\mum}{\mbox{$\mu^-$}}
\newcommand{\mup}{\mbox{$\mu^+$}}
\newcommand{\pp}{\mbox{$pp\rightarrow pp$}}
\newcommand{\pdHeh}{\mbox{$pd\to^3$He$\eta$}}
\newcommand{\pnpipipi}{\mbox{$pp\to pn\pi^{+}\pi^{+}\pi^{-}$}}

\begin{raggedright}

{\it A. Kupsc$^\dag$, P. Moskal$^\S$ and  M. Zieli\'nski$^\S$}\index{author}{ Kupsc, A.}\\
$^\dag$
Uppsala University, Uppsala, Sweden\\
$^\S$
Jagiellonian  University, Cracow, Poland
\bigskip


\end{raggedright}

\begin{center}
\textbf{Abstract}
\end{center}

Multimeson  production   in  $pp$  interactions   comprises  important
background  for   $\eta$,  $\omega$  and   $\eta'$  mesons  production
experiments  and for  the studies  of their  decays planned  with WASA
detector at  COSY.  The available  information about the  reactions is
summarized  and the  need for  efforts  to describe  the processes  is
stressed.

\section{Multimeson production}

Direct   production  of   three   or  more   pions  in   proton-proton
interactions,   has    not   received   proper    attention,   neither
experimentally nor  theoretically, despite the fact  that it comprises
the   main  background   for  \et,   \etp\  and   $\omega$  production
experiments.   With  a  4$\pi$  facility  such  as  WASA,  aiming  for
measurements  of  decays  of  $\eta$  and  $\eta'$  produced  in  $pp$
interactions  \cite{Adam:2004ch},  the  understanding of  the  \pptpi\
reactions  becomes   very  important  as  they   constitute  a  severe
background for studies of $\eta$  and $\eta'$ decays into three pions.
Those decays  provide key ingredients for determination  of the ratios
of light quark masses \cite{Gross:1979ur, Leutwyler:1996qg}, since the
partial  decay widths  are  proportional  to $d$  and  $u$ quark  mass
difference  squared.   In addition  precise  studies  of \etp\  decays
require     knowledge     of     $pp\to    pp\pi\pi\pi\pi$,     $pp\to
pp\pi\pi\pi\pi\pi$,   $pp\to  pp\pi\eta$   and   $pp\to  pp\pi\pi\eta$
reactions for beam energies  around $\eta'$ production threshold.  For
more  than  forty years  there  were  only  three experimental  points
available  for   the  cross  section  of   \pppipipi\  and  \pnpipipi\
reactions,    all    coming    from   bubble    chamber    experiments
\cite{PhysRev.125.2091,   PhysRev.126.747,   PhysRev.138.B670}.   Only
recently  the data  base  has  been extended  by  the measurements  of
\pppipipi\ and \pppopopo\ reactions  cross sections near the threshold
by  the  CELSIUS/WASA   collaboration  \cite{Pauly:2006pm}.   For  the
remaining reactions there is no data in that energy region.

The direct  production should proceed by  an excitation of  one or two
baryon    resonances    followed     by    the    subsequent    decays
\cite{PhysRev.123.333}.   For  example  in  the  case  of  three  pion
production  the  low  energy  region  a  mechanism  with  simultaneous
excitation of  $N^*$ and $\Delta(1232) P_{33}$  resonances is expected
to  dominate.  The  $N^*$ involved  has  to decay  into $N\pi\pi$  and
therefore  the  lowest lying  Roper  ($N(1440)  P_{11}$) and  $N(1520)
D_{13}$ resonances could be considered.

The influence of  the resonances can be studied  in the invariant mass
distributions  of the subsystems  of the  outgoing protons  and pions.
Such studies were done for  the \pppipipi\ and \pnpipipi\ reactions in
bubble chamber experiments performed  at higher energies (beam kinetic
energies  of  4.15  GeV and  9.11  GeV)  with  up to  thousand  events
\cite{Alexander:1967aa,  Colleraine:1967aa, Almeida:1969bv}.  However,
close to the threshold the analysis is not conclusive since the widths
of the  involved resonances are  comparable with the  available excess
energy ($Q)$.   Therefore, in this  case one expects that  phase space
distribution  and  the  final  state interaction  among  the  outgoing
nucleons  provide  a  reasonable  description of  the  observed  cross
sections.   The near  threshold cross  sections for  the  single meson
production  via the  nucleon-nucleon interaction  can indeed  be quite
satisfactory  described  by  such  ansatz.   For  the  productions  of
multiple  mesons the  assumption should  hold even  for  higher excess
energies since  on average  the energy available  to the pairs  of the
outgoing   particles  will   be  lower.    That  is   consistent  with
CELSIUS/WASA   results  on   $pp\to   pp\pi^+\pi^-\pi^0$  and   $pp\to
pp\pi^0\pi^0\pi^0$ reactions studied at $Q\approx 100$ MeV.

The production mechanism could  be studied instead by measuring ratios
of the cross sections for  the different charge states.  With the lack
of    information   the   simplest    assumption   is    Fermi   model
\cite{Fermi:1950jd}, where  amplitudes for all isospin  states are put
to be equal.  The assumption  leads to definite predictions for ratios
between  different  charge  states  of  the  reactions.   For  example
$\sigma(\pppipipi):    \sigma(\pppopopo):   \sigma(\pnpipipi)   \equiv
\sigma_1:\sigma_2:\sigma_3  =8:1:10$.  Resonances in  the intermediate
state will  modify the  ratios. The effect  can be illustrated  in the
Isobar  Model   \cite{PhysRev.123.333}  where  the   ratio  $\sigma_1:
\sigma_2:  \sigma_3$ is $7:1:25$  (assuming $\Delta  N^*$ intermediate
state)  or   $5:2:10$  (assuming  $N_1^*N_2^*$).    Experimentally:  $
\sigma_1: \sigma_3$  at beam kinetic energy  2.0 GeV (1:2.53$\pm$0.46)
\cite{PhysRev.125.2091}    and   at    2.85    GeV   (1:1.59$\pm$0.27)
\cite{PhysRev.126.747}.   The ratio $\sigma_1:  \sigma_2= 5.2\pm0.8:1$
was  measured at  lower  energy  -- 1.36  GeV  in recent  CELSIUS/WASA
experiment  \cite{Pauly:2006pm}.   One  trivial  modification  to  the
predicted  ratios close  to  threshold comes  from difference  between
volumes of  the phase spaces due to  $m_{\pi^{+}}\neq m_{\pi^{0}}$ and
it amounts to 18\% at 1.36 GeV.

Due to  lack of  microscopic model calculations,  of the same  kind as
those for single meson or for double pion production the semiclassical
Isobar  Model  is  at  present   the  only  option  to  describe  more
complicated  reactions  below  $\sqrt{s}\approx  5$ GeV.   The  modern
version  of the isobar  model is  used as  input to  relativistic ions
calculations    using     transport    equations    \cite{Teis:1996kx,
Bass:1998ca}.   Reliability  of  the  calculations can  be  tested  in
simpler cases by comparison  with existing calculations for the double
pion production \cite{AlvarezRuso:1997mx}.   The implementation of the
resonances in  the Isobar Model,  their production cross  sections and
decay branching  ratios can be  evaluated by studying  exclusive meson
production   reactions.    Multimeson   production  in   proton-proton
interactions  provides very  sensitive  test of  the parameters.   The
existing  calculations within  the framework  have focused  so  far on
production of dileptons in  proton-proton interactions with the aim to
understand  the background  for  high density  nuclear matter  probes
\cite{Faessler:2000md}.    The   byproduct   of  such   studies   were
calculations of the background for $\eta$ and $\eta'$ decays involving
dileptons (see  for example ref.~\cite{Faessler:2004fn}).   There is a
need  to  extend  the  calculations  to obtain  predictions  also  for
multimeson processes.

\section{Background for $\eta$, $\eta'$ decays}

A feature of  $\eta (\eta')$ detection from $pp\to  pp\eta (\eta')$ in
the WASA  detector is  the precise tagging  by missing  mass technique
with a resolution of a few MeV/c$^2$.  The resolution of the invariant
mass of the decay system is typically considerably worse.  Therefore a
figure of merit to describe background from direct production process,
leading to the identical final state as for an $\eta(\eta')$ decay, is
given by $\rho_B\equiv  d\sigma_B/d\mu|_{\mu=m_{\eta (\eta')}}$ -- the
differential cross  section for the  background at the  $\eta (\eta')$
peak  in  the $pp$  missing  mass  $\mu$.   Figure 1  shows  inclusive
$\rho_B$    values   for    $\eta'$   derived    from    the   COSY-11
measurements~\cite{Moskal:1998aa,Moskal:2000aa,
Khoukaz:2004si}.
\begin{figure}[ht]
\hspace{2.5cm}
\parbox{0.65\textwidth}{\vspace{-0.8cm}\centerline{
\includegraphics[width=0.69\textwidth]{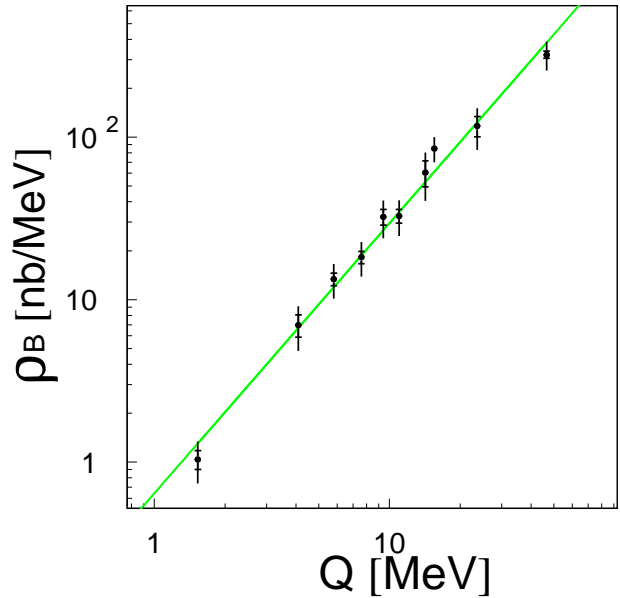}}}
\caption[aa]{ Inclusive differential cross section $\rho_B$ for
multipion production derived from the COSY-11 data~\cite{Moskal:1998aa,Moskal:2000aa,
Khoukaz:2004si}.  The
line is the parametrization $\rho_B= \alpha (Q/(1\
\mathrm{MeV}))^{\beta}$ fitted to the data: $\alpha = 0.64\pm 0.14\
[\mathrm{nb/MeV}]$ and $\beta = 1.66\pm 0.08$
\cite{Zielinski:2007aa}.  }
\end{figure}
 When estimating  the background for  a given decay  channel, quantity
$\rho_B\Delta\mu$ (where $\Delta \mu$ is the resolution in the missing
mass)  should be compared  to $\sigma_{\eta(\eta')}  BR_i$ (production
cross section times branching ratio for the decay mode).  The $\rho_B$
value  depends  on  the  total  cross  section  and  on  the  reaction
mechanism.   For  $\eta (\eta')$  production  close  to threshold  the
background distributions at the edge  of the phase space are relevant.
For both signal and the multimeson production this region of the phase
space is  strongly influenced by  $pp$ final state.   When approaching
the threshold  the $\rho_B$ decreases quickly and  in addition missing
mass resolution  improves (since  it is constrained  more by  the beam
momentum resolution) increasing signal to background ratio.  The price
is however a lower production cross section $\sigma_{\eta(\eta')}$.

Detailed  studies  of  $\eta\to\pi\pi\pi$  decays  in  $pp\to  pp\eta$
reaction at  beam energy 1.36 GeV by  CELSIUS/WASA collaboration shows
that background from direct three  pion production is 10--20\% for the
\po\pip\pim\  and 5\%  for the  \po\po\po\ channel  in the  final data
sample \cite{Pauly:2006pm, Bashkanov:2007iy}.  This allows for precise
study  of  the $\eta$  decays  providing  large  number of  events  is
collected.  For  $\eta'$ decays the situation is  quite different: the
three pion  decays have  branching ratios at  percent or  permil level
\cite{Yao:2006px}.   In  comparison  to  the $\eta$  meson  the  \etp\
production cross section at similar  excess energies is about 30 times
lower~\cite{Eta,Moskal:1998aa,Moskal:2000aa,
Khoukaz:2004si,Hibou:1998,Balestra:2000}.  Finally the  total cross section for multipion
reactions increases strongly when  going from \et\ to \etp\ production
threshold  region,  e.g.   for  \pptpi\ reaction  50--100  times.   In
conclusion embarking  on the \etp\ decay program  in $pp$ interactions
requires much better understanding of multimeson production reactions.



\end{document}

%% file: MENU07macros.tex
\newcommand{\beq}{\begin{equation}}
\newcommand{\eeq}[1]{\label{#1}\end{equation}}
\newcommand{\eeqn}{\end{equation}}

\newcommand{\beqa}{\begin{eqnarray}}
\newcommand{\eeqa}[1]{\label{#1}\end{eqnarray}}
\newcommand{\eeqan}{\end{eqnarray}}

\let\bar=\overbar

\newcommand{\Dslash}{\not{\hbox{\kern-4pt $D$}}}
\newcommand{\dslash}{\not{\hbox{\kern-2pt $\del$}}}

\newcommand{\msb}{{\bar{\ssstyle M \kern -1pt S}}}

